%Paper: hep-ph/9309211
%From: ULF MEISSNER <MEISSNER@ITP.unibe.ch>
%Date: Wed, 01 Sep 1993 16:05:04 MET

\magnification = 1200
\def\lapp{\hbox{$ {
\lower.40ex\hbox{$<$}
\atop \raise.20ex\hbox{$\sim$}
}
$}  }
\def\rapp{\hbox{$ {
\lower.40ex\hbox{$>$}
\atop \raise.20ex\hbox{$\sim$}
}
$}  }

\def\krig#1{\vbox{\ialign{\hfil##\hfil\crcr
$\raise0.3pt\hbox{$\scriptstyle \circ$}$\crcr\noalign
{\kern-0.02pt\nointerlineskip}
%{\kern-0.06pt\nointerlineskip}
$\displaystyle{#1}$\crcr}}}
\def\upar#1{\vbox{\ialign{\hfil##\hfil\crcr
$\raise0.3pt\hbox{$\scriptstyle \leftrightarrow$}$\crcr\noalign
{\kern-0.02pt\nointerlineskip}
$\displaystyle{#1}$\crcr}}}
\def\ular#1{\vbox{\ialign{\hfil##\hfil\crcr
$\raise0.3pt\hbox{$\scriptstyle \leftarrow$}$\crcr\noalign
{\kern-0.02pt\nointerlineskip}
$\displaystyle{#1}$\crcr}}}

\def\svec#1{\skew{-2}\vec#1}
\def\Tr{\,{\rm Tr }\,}

\def\g5{\gamma_5}

\def\lp1{{\cal L}_{\pi N}^{(1)}}
\def\lp2{{\cal L}_{\pi N}^{(2)}}
\def\lp3{{\cal L}_{\pi N}^{(3)}}

\topskip=0.60truein
\leftskip=0.18truein
\vsize=8.8truein
\hsize=6.5truein
\tolerance 10000
\hfuzz=20pt

\baselineskip 12pt plus 1pt minus 1pt
\pageno=0
\centerline{\bf CONSISTENT CALCULATION OF THE} \smallskip \centerline{\bf
NUCLEON ELECTROMAGNETIC POLARIZABILITIES}
\smallskip
\centerline{\bf IN CHIRAL PERTURBATION THEORY}
\smallskip  \centerline{{\bf  BEYOND NEXT-TO-LEADING ORDER
 }\footnote{*}{Work supported in part by Deutsche Forschungsgemeinschaft
and by Schweizerischer Nationalfonds.\smallskip}}
\vskip 24pt
\centerline{V\'{e}ronique Bernard}
\vskip 4pt
\centerline{\it Centre de Recherches Nucl\'{e}aires et Universit\'{e}
Louis Pasteur de Strasbourg}
\centerline{\it Physique Th\'{e}orique, Bat. 40A,
BP 20, 67037 Strasbourg Cedex 2,
France}
\vskip 12pt
\centerline{Norbert Kaiser, Armin Schmidt}
\vskip 4pt
\centerline{\it Physik Department T30,
Technische Universit\"at M\"unchen}
\centerline{\it
James
Franck Stra{\ss}e,
85747 Garching, Germany}
\vskip 12pt
\centerline{Ulf-G. Mei{\ss}ner\footnote{$^\dagger$}{Heisenberg
Fellow. Address after September 1$^{st}$, 1993: CRN, Physique Th\'eorique, BP
20, F-67037 Strasbourg Cedex 2, France.}} \vskip 4pt
\centerline{\it Universit\"at Bern,
Institut f\"ur Theoretische Physik}
\centerline{\it Sidlerstr. 5, CH--3012 Bern,\ \ Switzerland}
\vskip 0.5in
\centerline{\bf ABSTRACT}
\medskip
\noindent We calculate the nucleons' electromagnetic polarizabilities
in heavy baryon chiral perturbation theory including all terms to
order ${\cal O} (q^4)$. The chiral prediction of the electric polarizabilities
for the neutron and the proton are in good agreement with the data.
In the case of the magnetic polarizabilities the big positive
contribution from the $\Delta(1232)$ resonance is largely cancelled
by a non--analytic loop contribution of the $\ln M_\pi$ type.
This novel effect helps to understand the rather small empirical
value of the nucleons' magnetic polarizability.
\medskip
\vfill
\noindent BUTP--93/22  \hfill August 1993

\noindent CRN 93--39
\eject
\baselineskip 14pt plus 1pt minus 1pt
\noindent 1. The electric ($\bar \alpha$) and magnetic ($\bar \beta$)
polarizabilities of the proton and the neutron encode information
about the structure of these particles in the non--perturbative
regime of QCD.
Over the last few years, very precise experiments supplemented
with dispersion sum rules have established rather tight bounds on these
quantities. The outstanding features of these results are that the
proton and the neutron both behave  essentially as (induced)
electric dipoles
($\bar \alpha_p \simeq \bar \alpha_n \gg
  \bar \beta_p \simeq \bar \beta_n $)
and that the sums of electric and magnetic polarizabilities are equal
within ten percent,
$( \bar \alpha + \bar \beta)_p \simeq
( \bar \alpha + \bar \beta)_n $.
The first of these observations appears rather stunning since the
very strong magnetic $N \Delta$
transition gives rise to
a large positive contribution to the magnetic polarizabilities.
Therefore, it has been
argued by many that a theoretical explanation of these fundamental quantities
has to involve explicit $\Delta(1232)$ degrees of freedom.
This theme
will be picked up below. On the theoretical side, chiral perturbation theory
(CHPT) offers the best venue of understanding these structure constants.
Indeed, in refs.[1,2]  it was shown that the nucleon electromagnetic
polarizabilities to leading order in the chiral expansion are pure
one--loop effects. Consequently, at this order they can be calculated
without any unknown parameters. This is similar to the prediction for the
charge radius of the neutral kaon or other mesonic
processes like $\gamma \gamma \to \pi^0 \pi^0$, $K_L \to \pi^0 \gamma \gamma$
and $K_S \to \gamma \gamma$.
In refs.[1,2], the
nucleon was treated  as a fully relativistic Dirac field. The fact that the
nucleon mass  does not vanish in the chiral limit poses some problems to the
chiral power counting in the effective meson-nucleon Lagrangian. In particular,
the one loop result leads to a string of terms in increasing powers of the
pion mass. From these, only the first one is not affected by contributions
from  higher loop orders not yet calculated.
In fact, the calculation of refs.[1,2]
was redone in ref.[3] in the framework of heavy baryon CHPT (HBCHPT),
which allows for a consistent chiral power counting, i.e. loops are
suppressed by powers of $q^2$ with $q$ a genuine small four--momentum
 (this is not the case if one treats the nucleon fully relativistically).
In the framework of HBCHPT, only the leading singular term proportional
to $1/M_\pi$ in the electromagnetic polarizabilities survives and one
finds a surprisingly good description of the data (see below). However,
as mandated by the decoupling theorem [4], in this calculation
the intermediate states are only nucleons, i.e. no $\Delta$'s appear.
As already mentioned, the absence of the low--lying spin--3/2
resonances in this approach has prompted many critical remarks concerning the
results to leading order ${\cal O}(q^3)$. Jenkins and Manohar [5] have
argued that it is mandatory to include the spin-3/2 decuplet in the
effective theory to cancel large SU(3) loop effects. In this spirit,
Butler and Savage [6]  have performed a calculation of the polarizabilities
in SU(3)\footnote{$^{1)}$}{Notice that the first SU(3) calculation of
these observables was performed in ref.[7].} with explicit decuplet
degrees of freedom.
However, the whole approach suffers
from two shortcomings. First, including the decuplet destroys the consistent
power counting of HBCHPT. This can be traced back to the non--vanishing
scale related to the finite decuplet--octet mass splitting which does
not vanish in the chiral limit [8].\footnote{$^{2)}$}{This might be
circumvented if one considers the $N \Delta$ mass splitting as a small
parameter. However, in three-color QCD this quantity can not be considered
arbitrarily small in contrast to the light quark masses or external momenta of
the fundamental fields in the effective field theory.}
Furthermore, the
calculations performed in refs.[5,6] do not include all terms in a given order
in CHPT but only some non--analytic loop contributions and some
analytic terms. Clearly, this does not conform to the philosophy underlying
CHPT, namely to include {\it all} terms at a given order.
The $\Delta(1232)$ was also hailed as a deus ex machina in explaining
the data of the nucleon electromagnetic polarizabilities by L'vov [9].
In this paper, we will present a complete CHPT calculation of these
fundamental quantities including all effects up to (and including) order $q^4$
instead of arguing about this or that being the dominant effect.
\bigskip
\noindent 2.
Consider the spin--averaged
Compton amplitude for scattering off nucleons at low energies.
The energy expansion of the spin--averaged
Compton amplitude in the nucleon rest
frame takes the form
$$T(\gamma N \to \gamma N) = - { e^2 Z^2 \over 4 \pi m}
+ \bar \alpha \, \omega'\omega \, \svec
\epsilon'\cdot \svec \epsilon + \bar \beta\,
( \svec \epsilon'\times
\svec k')\cdot ( \svec \epsilon \times \svec k) + {\cal O}(\omega^4)
\eqno(1)$$
with ($\omega, \svec k, \svec \epsilon$) and
($\omega', \svec k', \svec \epsilon'$) the frequencies, momenta and
polarization vectors of the incoming and outgoing photon, respectively.
The first term, which is energy--independent, is nothing but the
Thomson scattering amplitude as mandated by gauge invariance. It
constitutes a low--energy theorem as the photon energy tends to zero.
At next--to--leading order in the energy expansion, the photon probes
the non--trivial structure of the spin--1/2 particle (here, the
nucleon) it scatters off. This information is encoded in two structure
constants, the so--called electric ($\bar \alpha$) and magnetic
($\bar \beta$) polarizabilities. These have been determined rather
accurately over the last years. For the proton, if one
combines the
Illinois, Mainz and Saskatoon measurements [10] with the dispersion sum rule,
$(\bar \alpha+ \bar \beta)_p = 14.2 \pm 0.3$ [11], one has
$$
\bar \alpha_p
= 10.4 \pm 0.6
\, , \, \, \,
\bar \beta_p   = 3.8 \mp 0.6   \eqno(2)$$
all in units of $10^{-4}$ fm$^3$ which we will use throughout and do not
exhibit explicitely any further. Notice that we have added the statistical and
systematic errors in quadrature. Similarly, the dispersion sum rule
$(\bar \alpha+ \bar \beta)_n = 15.8 \pm 0.5$ [12] \footnote{$^{3)}$}{Notice
that
the error on this number is presumably underestimated since one has to use
deuteron data to extract the photon-neutron cross section.}
together with  the recent Oak Ridge and Mainz [13] measurements lead to
$$\bar \alpha_n   = 12.3  \pm 1.3 \, , \, \, \,
\bar \beta_n   = 3.5 \mp 1.3  \eqno(3)$$
The salient features of these experimental results have already been
mentioned, namely that both the proton and the neutron essentially behave as
(induced)
electric dipoles and that their respective sums of electric and magnetic
polarizabilities are almost the same. Let us stress again that naively one
expects a large contribution from the $\Delta(1232)$ to the magnetic
polarizabilities (of the order of 10) and it thus appears difficult to explain
the relatively small values of $\bar{\beta}_{p,n}$.
\bigskip
\noindent 3. To calculate the polarizabilities, we make use of HBCHPT.
 We work in flavor SU(2) and in the isospin limit
$m_u = m_d         $. The relevant degrees of freedom of the effective
Lagrangian are the pions and the nucleon. The Goldstone fields are collected
in the matrix
$U(x) = \exp[i{\vec \tau}\cdot{\vec \pi}(x) / F_\pi] = u^2 (x)$ and the
nucleons are considered as very heavy, i.e. non--relativistically [14]. The
effective Lagrangian to order ${\cal
O}(q^4)$, where $q$ denotes a genuine small momentum or a meson
(quark) mass, reads
(we only exhibit those terms which are actually needed in the calculation)
$$\eqalign{{\cal L}_{\pi N} &= {\cal L}_{\pi N}^{(1)} +
{\cal L}_{\pi N}^{(2)} + {\cal L}_{\pi N}^{(4)} \cr
{\cal L}_{\pi N}^{(1)} &= {\bar H} ( i v\cdot D + g_A S \cdot u ) H \cr
{\cal L}_{\pi N}^{(2)} &= {\bar H} \biggl\lbrace  -{1\over 2m} D \cdot D +
{1 \over 2m} (v \cdot D)^2 +   c_1   \Tr \chi_+             + \bigl( c_2 -
{g_A^2
\over 8 m} \bigr)  v\cdot u\, v\cdot u \cr & + c_3\,  u\cdot u - {i g_A \over
2m} \lbrace S \cdot D , v \cdot u \rbrace - { i e \over 4 m}
[ S^\mu , S^\nu ] \biggl( (1 + \kappa_v) f^+_{\mu \nu} + {\kappa_s - \kappa_v
        \over 2 } \Tr f^+_{\mu \nu} \biggr)
\biggr\rbrace H \cr
{\cal L}_{\pi N}^{(4)} &= {\pi \over 4} (\delta \bar \beta_p -
\delta \bar \beta_n) \bar H f_{\mu \nu}^+ f^{\mu \nu}_+ H
                        + {\pi \over 4} \delta \bar \beta_n
                        \bar H  H \Tr f_{\mu \nu}^+ f^{\mu \nu}_+  \cr
                       &+ {\pi \over 2} (\delta \bar \alpha_n
 + \delta \bar \beta_n
 - \delta \bar \alpha_p
 - \delta \bar \beta_p)
    \bar H f_{\mu \nu}^+ f^{\lambda \nu}_+ H v^\mu v_\lambda  \cr
                       &- {\pi \over 2} (\delta \bar \alpha_n
 + \delta \bar \beta_n)
    \bar H H \Tr(f_{\mu \nu}^+ f^{\lambda \nu}_+ ) v^\mu v_\lambda
 \cr} \eqno(4)$$
with
$$\eqalign{u_\mu &= i u^\dagger \nabla_\mu U u^\dagger \cr
f_{\mu \nu}^+ &= (\partial_\mu A_\nu - \partial_\nu A_\mu)
(u Zu^\dagger + u^\dagger Zu)
\cr} \eqno(5)$$
where $H$ denotes the heavy nucleon field of charge $Z = (1+ \tau_3)/2$
and anomalous
magnetic moment $\kappa = (\kappa_s + \tau_3 \kappa_v)/2$,
$S_\mu$ the covariant spin--operator
subject to the constraint $v \cdot S = 0$, $\nabla_\mu$ the covariant
derivative acting on the pions and we adhere to the notations of ref.[3]. The
superscripts (1,2,4) denote the chiral power.     The lowest order effective
Lagrangian is of order ${\cal O}(q)$. The one loop contribution is suppressed
with respect to the tree level by $q^2$ thus contributing at ${\cal O}(q^3)$.
In addition, there are one loop diagrams with exactly one insertion from
${\cal L}_{\pi N}^{(2)}$. These are of order $q^4$.
Finally, there are contact terms of order $q^2$ and $q^4$ with
coefficients
not fixed by chiral symmetry. For the case at hand, ${\cal L}_{\pi N}^{(3)}$
does not contribute. Notice that  some
coefficients in ${\cal L}_{\pi N}^{(2)}$ related to the $\gamma \gamma N N$
and $\gamma \pi N N$ vertices are fixed from the relativistic theory
by the low-energy theorems for Compton
scattering and neutral
pion photoproduction, respectively. This is discussed in some
detail in ref.[3]. The unknown coefficients we have to determine are $c_1$,
$c_2$ and $c_3$ as well as the four low-energy constants $\delta
\bar{\alpha}_p$, $\delta \bar{\alpha}_n$, $\delta \bar{\beta}_p$ and $\delta
\bar{\beta}_n$ from ${\cal L}_{\pi N}^{(4)}$.
We have not exhibited the standard meson Lagrangian ${\cal L}_{\pi \pi}^{(2)}$.

To calculate all terms    up-to-and-including order $q^4$, we have to consider
all
one loop graphs with insertions from ${\cal L}_{\pi N}^{(1)}$ and those
with exactly one
insertion from ${\cal L}_{\pi N}^{(2)}$. While the former scale as  $q^3$, the
latter constitute the
new contributions of order $q^4$. Furthermore,     there are the
tree diagrams related to ${\cal L}_{\pi N}^{(4)}$ which are also new. It is
worth to stress that in the one loop diagrams involving ${\cal L}_{\pi
N}^{(2)}$ the anomalous magnetic moment of the nucleon appears.
Finally, it is
mandatory to expand the leading order ${\cal O}(q)$ effective vertices
as well as the
nucleon propagator in momenta
to include all relativistic
corrections of order
$1 / m$. The details of these calculations can be found in ref.[15].
For further convenience, we introduce the constant $C$,
$$ C = {e^2   \over 96 \pi^2 F_\pi^2}\, \, = 4.36 \cdot 10^{-4} \, {\rm fm}^2
\eqno(6)$$
with $e^2 / 4 \pi = 1/ 137.036$ and $F_\pi \simeq 93$ MeV the weak pion decay
constant. The resulting expressions for the electric polarizabilities of the
proton and the neutron can most compactly be written as
$$\bar{\alpha}_i = {5 C g_A^2 \over 4 M_\pi} + {C \over \pi} \biggl[ \bigl(
{x_i
g_A^2 \over m} - c_2 \bigr) \ln {M_\pi \over \lambda} + {1 \over 4} \bigl(
{y_i g_A^2
\over 2 m} - 6 c_2 + c^+ \bigr) \biggr] +\delta
\bar{\alpha}_i^r(\lambda) \, , \quad i =
p, \, n \eqno(7)$$ with
$$x_p = 9 \, , \quad x_n = 3 \, , \quad y_p = 71 \, , \quad y_n = 39 \, .
\eqno(7a)$$
Here,  $\lambda$ is the scale of dimensional regularization. The
physical $\bar \alpha_i$ are of course independent of this scale since
the renormalized counter terms $\delta \bar \alpha_i^r(\lambda)$
cancel  the logarithmic $\lambda$--dependence of the loop contribution.
The first term on
the r.h.s. of eq.(7) is, of course, the result at order $q^3$ [3]. Similarly,
one finds for the corresponding magnetic polarizabilities
$$\bar{\beta}_i = { C g_A^2 \over  8 M_\pi} + {C \over \pi}  \biggl[ \bigl( {3
x_i' g_A^2 \over m} - c_2 \bigr) \ln {M_\pi \over \lambda} + {1 \over 4}
\bigl( {y_i' g_A^2
\over  m} + 2 c_2 - c^+ \bigr) \biggr] + \delta \bar{\beta}_i^r (\lambda)
\, , \quad i =
p, \, n \eqno(8)$$ with
$$x_p' = 3 + \kappa_s \, , \quad x_n' = 1 - \kappa_s \, , \quad y_p' = {37
\over
2} + 6 \kappa_s \, , \quad y_n' = {13 \over 2} - 6 \kappa_s \, . \eqno(8a)$$
We have introduced the coefficient $c^+$ related to $c_1$, $c_2$ and $c_3$,
$$c^+ = -8 c_1 + 4 c_2 + 4 c_3 - {g_A^2 \over 2 m} \, \, \, . \eqno(9)$$
The results shown in eqs.(7,8) have the following structure.
Besides the $1/M_\pi$ term, which is discussed in detail in refs.[2,3], the
${\cal O}(q^4)$ contributions from the loops have a $\ln M_\pi$ and a constant
piece $\sim M_\pi^0$. As a check one can recover the coefficient of the
$\ln M_\pi$ term form the relativistic calculation [1,2] if one sets
the new low energy constants $c_i$ and $\kappa_{s,v} = 0$. In that
case only the $1/m$ corrections of the relativistic Dirac formulation
are treated and one necessarily reproduces the corresponding
non--analytic (logarithmic) term of this approach.
The term proportional to $c_2 \, \ln M_\pi$ in eqs.(7,8) represents the effect
of (pion) loops with intermediate $\Delta (1232)$ states [6] consistently
truncated at order $q^4$.
Notice that from now on we will omit the superscript '$r$' on
$\delta \bar \alpha_i^r$ and
$\delta \bar \beta_i^r$ appearing in
eqs.(7,8).
Finally,  from eqs.(7) and (8) one can form two quantities which are
independent of the low-energy constants $c_2$ and $c^+$ and which vanish to
leading order $q^3$,
$$\eqalign{
\Sigma^+ &= (\bar{\alpha} + \bar{\beta})_p - (\bar{\alpha} + \bar{\beta})_n
\cr
&= {C \over \pi}{g_A^2\over m} \biggl[ 6(2+\kappa_s) \ln{M_\pi \over \lambda} +
(7 + 3\kappa_s) \biggr] + \delta \bar{\alpha}_p(\lambda)
+ \delta \bar{\beta}_p(\lambda) -
\delta \bar{\alpha}_n(\lambda) - \delta \bar{\beta}_n(\lambda)
\cr
\Sigma^- &= (\bar{\alpha} - \bar{\beta})_p - (\bar{\alpha} - \bar{\beta})_n
\cr
&= {C \over \pi}{g_A^2\over m} \biggl[1- 3\kappa_s(1+ 2 \ln{M_\pi \over
\lambda})
\biggr] + \delta \bar{\alpha}_p(\lambda) - \delta \bar{\beta}_p(\lambda)
- \delta \bar{\alpha}_n(\lambda) + \delta \bar{\beta}_n(\lambda)
\cr} \eqno(10) $$
Empirically, one has $\Sigma^+ = -1.6$ and $\Sigma^- = -2.2$. From these
numbers one immediately realizes the importance of isospin breaking in
the counter terms from
${\cal L}_{\pi N}^{(4)}$. Setting their renormalized values equal for
the proton and the neutron
and choosing $\lambda = 1.232$
GeV, one finds $\Sigma^+ = -8.3$ which deviates considerably from the
empirical value. Similarly, the prediction $\Sigma^- = -0.1$ is not in good
agreement with the data. We will come back to these quantities later on.
\bigskip
\noindent 4. We now have to determine the values of the various low-energy
constants appearing in eqs.(7) and (8). The constants $c_{1,2,3}$ already
appeared in the study of the chiral corrections to the S--wave pion--nucleon
scattering lengths [16]. From that paper, we can deduce the relation
$$c^+ = {256 \pi^2 F_\pi^4 a^+ - 3 g_A^2 M_\pi^3 \over 32 \pi M_\pi^2 F_\pi^2
( 1 - M_\pi /m )} = - 0.33 \, {\rm fm}
\eqno(11)$$
where $a^+$ is the isospin--even S-wave scattering length, $a^+ = -0.83 \pm
0.38 \cdot 10^{-2} / M_\pi$, $g_A$ the axial-vector coupling constant
and $M_\pi = 139.57$ MeV the (charged) pion mass. Inserting the value of
$c^+$ as given from eq.(10) into eqs.(7,8), one finds that the contributions
proportional to $c^+$ to the electromagnetic polarizabilities are negligible
since they are less then 0.12 in magnitude. To determine $c_2$,
we make use of the principle of resonance saturation [17].
It states that to a high degree of accuracy the low--energy constants can be
calculated from resonance exchanges by integrating out the heavy resonance
fields from an
effective Lagrangian of the pions chirally coupled to the various resonances.
In the meson sector, this has been shown to work very well. We extend this
method to the baryon sector since it is essentially the only method of
estimating the unknown coefficients. For the constant $c_2$ we find
$$c_2 = {g_A^2 \over 2 ( m_\Delta -m )} = 0.59 \, {\rm fm}   \eqno(12)$$
where we have treated the $\Delta$ non--relativistically (isobar model)
and neglected a small contribution
from the Roper resonance. We use $m_\Delta = 1232$ MeV and $m = 938.27$ MeV.
A detailed discussion on the admissible range for $c_2$ including also  its
dependence on the off-shell parameter Z if the $\Delta$ is treated
relativistically (as a Rarita--Schwinger spinor) can
be found in ref.[15]. Here, we will use the value given in eq.(12) at
the scale $\lambda =m_\Delta$ according to the resonance saturation
principle.

The $\Delta(1232)$ enters prominently in the determination of the four
low-energy constants  from ${\cal L}_{\pi N}^{(4)}$.
Therefore, we will determine these coefficients at the scale
$\lambda = m_\Delta$.
In particular, one gets
a sizeable contribution to the magnetic polarizabilties due to the strong
$N\Delta$ M1 transition.
A crude estimate of this has been given in ref.[18]
by integrating the M1 part of the total photoproduction cross section for
single pion photoproduction over the resonance region,
$$\delta \bar{\beta}_p^\Delta (m_\Delta) = {1 \over 2 \pi^2} \int {d \omega
\over \omega^2} \, \sigma^{M1}(\omega) = 7.0  \eqno(13)$$
However, there is a large uncertainty this number. If one simply uses the Born
diagrams with an intermediate point--like
$\Delta$, one has
$\delta \bar{\beta}_p^\Delta (m_\Delta) = 10 \ldots 12$ [18] where the
uncertainty stems from the uncertainties in the $\gamma N \Delta$ coupling
strength. It has been argued in ref.[19] that an off-shell effect could
reduce this
number by almost an order of magnitude, however, this is outside the accuracy
of the chiral expansion performed here. Furthermore, from isospin arguments
one can conclude that $\delta \bar{\beta}_p^\Delta (m_\Delta) =
\delta \bar{\beta}_n^\Delta (m_\Delta) \simeq 7$. In addition to this
resonance contribution, one has also an effect from the (charged)
kaon loops [7]. Since
we are working in SU(2), the kaons and etas are frozen out and effectively
give some finite contact terms. One has for the electric polarizabilities
$$\eqalign{
\delta \bar{\alpha}_p^K (m_\Delta) &= {5  C \over 4}
{2D^2/3+2F^2 \over M_K } = 2.0 \cr
\delta \bar{\alpha}_n^K (m_\Delta) &= {5 C \over 4}
{(D-F)^2 \over M_K } = 0.2 \cr} \eqno(14)$$
for $D = 0.8$ and $F = 0.5$.
The corresponding numbers for the kaon contributions to the magnetic
polarizabilities are a factor 10 smaller [7]. With this, we have determined
all the unknown coefficients and are at the position of evaluating eqs.(7,8).
A more detailed discussion of these issues can be found in ref.[15].
\bigskip
\noindent 5. We now present our numerical results. We always use the
Goldberger-Treiman relation to express $g_A / F_\pi$ as $g_{\pi N} / m$ ,
with $g_{\pi N} = 13.40$ the strong pion--nucleon coupling constant.
We find with $\kappa_s = -0.12$
$$\eqalign{
\bar{\alpha}_p = 13.6 - 8.3 + 3.2 + 2.0 = 10.5 \cr
\bar{\alpha}_n = 13.6 - 1.6 + 1.2 + 0.2 = 13.4 \cr
\bar{\beta}_p  = 1.4  - 7.9 + 2.8 + 7.2 = 3.5  \cr
\bar{\beta}_n  = 1.4  - 2.0 + 1.4 + 7.0 = 7.8  \cr}
\eqno(15)$$
where we have separated the $q^3$ contribution (first term) from the $q^4$
ones which are of the $\ln M_\pi$, $M_\pi^0$ and counter
term type. Several
remarks on these results are in order. For the electric polarizabilities, we
find small corrections which bring the chiral prediction in agreement with the
data, in particular the difference in the proton to neutron electric
polarizabilities.  For the magnetic polarizabilities, the most important
result is that the large contribution from the $\Delta(1232)$ which enters
through the counter terms is mostly cancelled by the large coefficient of the
$\ln M_\pi$ term in case of the proton. For the neutron, this term is smaller
and therefore the resulting chiral prediction is too large. Clearly, to
overcome this at order $q^4$, one would need some substantial isospin breaking
in the photon-nucleon-Delta coupling constants (for which at present we have
no indication). Consequently, the quantity $\Sigma^+$ defined in eq.(10) comes
out too big, $\Sigma^+ = -7.2$. Also, we find $\Sigma^- = 1.5$ which has the
wrong sign due to the to large magnetic polarizability of the neutron. Is it
evident at this stage that a better understanding of the counter
terms from
${\cal L}_{\pi N}^{(4)}$ is necessary to further tighten the chiral prediction
for the magnetic polarizabilities of the proton and the neutron.
\bigskip \noindent
6. To summarize, we have used heavy baryon chiral perturbation theory to
calculate all corrections up--to--and--including order $q^4$ to the electric
and magnetic polarizabilities of the proton and the neutron.
To estimate the strength of the various
contact terms, we have made to some extent use of the principle of resonance
saturation which
is known to work very accurately in the meson sector. The main results of this
investigation are:
\medskip
\item{$\bullet$}{The corrections to order $q^4$ bring the chiral prediction of
the proton and neutron electric polarizabilities in agreement with the data,
$\bar{\alpha}_p = 10.5 \cdot 10^{-4}$ fm$^3$ and
$\bar{\alpha}_n = 13.4 \cdot 10^{-4}$ fm$^3$. The corrections to the leading
order ($q^3$) result $\bar{\alpha}_p = \bar{\alpha}_n = 13.6 \cdot 10^{-4}$
fm$^3$
are small and one thus expects no drastic changes to these results from the
next order, ${\cal O}(q^5)$.}
\medskip
\item{$\bullet$}{The situation is different for the magnetic polarizabilities.
There is undoubtedly a large positive contribution from the $\Delta(1232)$
resonance. Its actual value, however, is quite uncertain. For the proton, one
finds a substantial cancellation between the positive $\Delta$ and large
negative $\ln M_\pi$ term bringing the chiral prediction in good agreement
with the data. For the neutron, the $\ln M_\pi$ term has a considerably
smaller coefficient and thus the resulting value for $\bar{\beta}_n$ is
approximately a factor two too large.}
\medskip    \noindent
A more detailed account of these topics, in particular a discussion of the
allowed ranges for the various counterterms (which was not addressed in detail
here) will be presented in ref.[15]. Finally, it is worth to stress again that
the calculation presented here contains all effects up-to-and-including order
$q^4$ in the chiral expansion of the nucleons' electromagnetic
polarizabilities and that the resulting numbers reflect most of the (although
not all) empirical facts about these fundamental quantities.
\bigskip
\bigskip  \noindent
We are grateful to Alan Nathan, Nimai Mukhopadhyay and Ron Workman for
instructive comments.
\vfill \eject
\noindent
{\bf References}
\bigskip
\item{1.}V. Bernard, N. Kaiser and Ulf-G. Mei{\ss}ner,
{\it Phys. Rev. Lett.\/}
{\bf 67} (1991) 1515.
\smallskip
\item{2.}V. Bernard, N. Kaiser, and Ulf-G. Mei{\ss}ner, {\it Nucl. Phys.\/}
{\bf B373} (1992) 364.
\smallskip
\item{3.}V. Bernard, N. Kaiser, J. Kambor
and Ulf-G. Mei{\ss}ner, {\it Nucl. Phys.\/} {\bf B388} (1992) 315.
\smallskip
\item{4.}J. Gasser and A. Zepeda, {\it Nucl. Phys.\/}
{\bf B174} (1980) 445.
\smallskip
\item{5.}E. Jenkins and A.V. Manohar, {\it Phys. Lett.\/} {\bf B259} (1991)
353.
\smallskip
\item{6.}M. N. Butler and M. J. Savage, {\it Phys. Lett.} {\bf B294} (1992)
369.
\smallskip
\item{7.}V. Bernard, N. Kaiser, J. Kambor and Ulf-G. Mei{\ss}ner,
{\it Phys. Rev.\/}
{\bf D46} (1992) 2756.
\smallskip
\item{8.}V. Bernard, N. Kaiser and Ulf-G. Mei{\ss}ner,
{\it Z. Phys.\/}
{\bf C} (1993) in print.
\smallskip
\item{9.}A. L'vov, {\it Phys. Lett.} {\bf B304} (1993) 29.
\smallskip
\item{10.}A. Zieger {\it et al.}, {\it Phys. Lett.\/} {\bf B278} 34.

F.J. Federspiel {\it et al.}, {\it Phys. Rev. Lett.\/} {\bf 67}
(1991) 1511;

E.L. Hallin {\it et al.}, "Compton Scattering From The Proton", Saskatoon
preprint,

1993.
\smallskip
\item{11.}M. Damashek and F. Gilman, {\it Phys. Rev.\/} {\bf D1} (1970) 1319.
\smallskip
\item{12.}V.A. Petrunkin, {\it Sov. J. Nucl. Phys.\/} {\bf 12} (1981) 278.
\smallskip
\item{13.}K.W. Rose {\it et al.}, {\it Phys. Lett.\/} {\bf B234}
(1990) 460;

J. Schmiedmayer {\it et al.}, {\it Phys. Rev. Lett.\/} {\bf 66}
(1991) 1015.
\smallskip
\item{14.}E. Jenkins and A.V. Manohar, {\it Phys. Lett.\/} {\bf B255} (1991)
558.
\smallskip
\item{15.}A. Schmidt, V. Bernard, N. Kaiser and Ulf-G. Mei{\ss}ner,
in preparation.
\smallskip
\item{16.}V. Bernard, N. Kaiser and Ulf-G. Mei{\ss}ner,
{\it Phys. Lett.\/}
{\bf B309} (1993) 421.
\smallskip
\item{17.}G. Ecker, J. Gasser, A. Pich and E. de Rafael,
{\it Nucl. Phys.\/} {\bf B321} (1989) 311.
\smallskip
\item{18.}N.C. Mukhopadhyay, A.M. Nathan and L. Zhang,
{\it Phys. Rev.\/} {\bf D47} (1993) R7.
\smallskip
\item{19.}N.M. Butler, M.J. Savage and R. Springer,
{\it Nucl. Phys.\/} {\bf B399} (1993) 69.
\smallskip
%\item{xx.}R.D. Peccei, {\it Phys. Rev.\/} {\bf
%176} (1968) 1812.
%\smallskip
%\item{xx.}
%Ulf-G. Mei{\ss}ner, "Recent Developments in Chiral Perturbation Theory", Bern
%University preprint BUTP-93/01, 1993, to appear in {\it Rep. Prog.
%Phys.}$\,$. \smallskip
%\smallskip
\vfill \eject \end